\documentclass[openacc]{rstransa}

\newcommand{\apj}{Astrophys. J. }

\newcommand{\aap}{Astron. Astrophys.}
\newcommand{\apjl}{Astrophys. J. }
\newcommand{\mnras}{Mon. Not. R. Astron. Soc.}

\newcommand{\nat}{{\it Nature}}

\newcommand{\araa}{ARA\&A}

\newcommand{\ssr}{SSR}

\newcommand{\prl}{Physical Review Letters}

\def\araa{{Annual Reviews of Astronomy and Astrophysics}}   

\titlehead{Research}

\begin{document}

\title{Gravitational lensing in gamma-ray bursts}

\author{
A.J.\ Levan$^{1,2}$\!, B.P.\ Gompertz$^{3}$\!, G.P.\ Smith$^{3,4}$\!\!, M.E.\ Ravasio$^{1}$\!, G.P.\ Lamb$^{5}$\!, N.R.\ Tanvir$^{6}$}

\address{$^{1}$Department of Astrophysics/IMAPP, Radboud University Nijmegen;
$^{2}$Department of Physics, University of Warwick;
$^{3}$ School of Physics and Astronomy, University of Birmingham;
$^4$ Department of Astrophysics, University of Vienna;
$^{5}$ Astrophysics Research Institute, Liverpool John Moores University;
$^{6}$ Department of Physics and Astronomy, University of Leicester
}
\subject{xxxxx, xxxxx, xxxx}


\corres{Andrew Levan\\
\email{a.levan@astro.ru.nl}}

\begin{abstract}
Gravitationally lensed Gamma-ray bursts (GRBs) offer critical advantages over other lensed sources. They can be detected via continuously operating detectors covering most of the sky. They offer extremely high time resolution to determine lensing delays and find short-time delays accurately. They are detectable across most of the visible Universe. However, they are also rare and frequently poorly localized. In this paper, we review searches for gravitational lensing in GRBs and comment on promising avenues for the future. We note that the highly structured jets in GRBs can show variations on sufficiently small scales that, unlike lensing of most transient sources, gravitational lensing in gamma-ray bursts may not be achromatic. Such behavior would weaken the stringent requirements for identifying lensed bursts but would also make robust identification of lensing more challenging. A continuously running search that could identify candidate lensed events in near real-time would enable afterglow searches with current and near-future wide-field optical/IR surveys that could yield the first unambiguous detection of a lensed GRB. The new generation of sensitive X-ray and gamma-ray detectors, such as the Einstein Probe and SVOM, will complement {\em Swift} and significantly enhance the number of well-localized gamma-ray and X-ray transients. Tuned strategies could dramatically improve the probability of observing a lensed GRB.
\end{abstract}


\begin{fmtext}








\end{fmtext}


\maketitle

\section{Introduction}
Gamma-ray bursts (GRBs) have been known for more than 50 years \cite{klebesadel73} and manifest as flashes of high-energy radiation from cosmological distances \cite{metzger97}, which exhibit durations from milliseconds to at least several hours \cite{levan14}. They arise from at least two distinct progenitor channels, the collapse of massive stars  \cite{hjorth03,stanek03} or the merger of two compact objects \cite{tanvir13}, with the latter observed in one case in co-incidence with a gravitational wave signal \cite{abbott_bns}. Traditionally, the progenitors have been assigned to the two apparent duration classes \cite{kouveliotou93} with short-GRBs (durations $<2$s) thought to arise from compact object mergers, and long-GRBs (durations $>2$s) from the core collapse of massive stars \cite{levan16}. However, recent observations paint a more nuanced picture with several long GRBs showing signatures consistent with an origin in compact object mergers \cite{rastinejad22,troja22,levan24,yang24}. Collapsar-driven GRBs (sometimes referred to as type II events \cite{zhang09}) have been observed to at least $z \sim 8$ \cite{tanvir09,cucchiaria11}, while merger-driven events (also sometimes type I) are restricted to typically lower redshifts, but have also been observed at $z>2$ \cite{fong22}. Indeed, the median redshift of GRBs is around $z \sim 2$ \cite{jakobsson06}, much larger than for extragalactic transients detected at optical wavelengths. GRB energy budgets are extreme at these distances, with energy, if released isotropically exceeding $10^{55}$ erg in some cases \cite{malesani23,burns23}. However, beaming likely reduces the typical ``true" energy to $\sim 10^{51}$ erg \cite{frail01,berger03}. 

The combination of high luminosity and subsequent large distances makes GRBs appealing prospects for gravitational lensing \cite{porciani01}. GRBs will occasionally have lines of sight which pass sufficiently close to foreground masses to be gravitationally lensed and so will appear to detectors as two (near) identical GRBs in terms of their temporal profiles, with 
overlapping spatial localisation (even the best gamma-ray detectors to date only provide positions accurate to the arcminute level, although afterglows frequently provide sub-arcsecond positions).
The time separations could be as short as milliseconds but up to timescales of months or years, depending on the lens mass and structure and the geometry of the situation. Such lensed GRBs provide many of the same diagnostics available via the study of other time-variable lensed objects such as supernovae \cite{refsdal64,kelly15,rodney21} or quasars \cite{suyu10}, but also can probe much shorter delays, and offer a route to new tests of very low mass lenses and consequent short time delays created by them. Since quasars and supernovae vary much more slowly, the field of short-delay lens systems is unexplored from an observational perspective.

This article outlines the current state of GRB lensing, its advantages and disadvantages,  previous searches and prospects for future lensing observations.

\section{The advantages of GRB lensing} 
\begin{figure*}
	\centering
		\includegraphics[scale=0.6,angle=0]{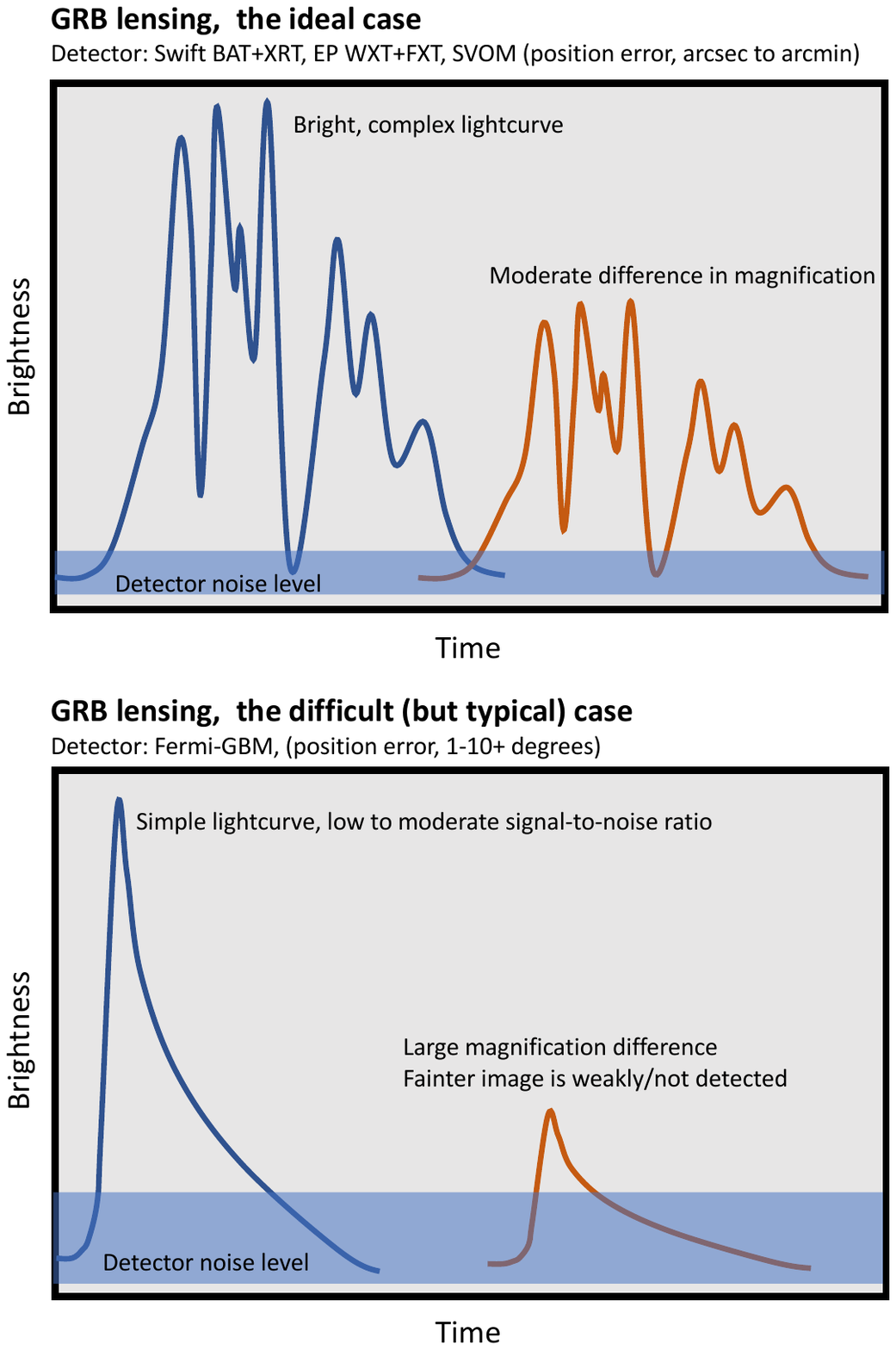}
	\caption{A cartoon of different scenarios for GRB lensing. The upper panel shows an ideal scenario where lensing could be claimed purely on the basis of the similarity of the $\gamma$-ray lightcurves (and spectra) due to the intrinsic complexity of the lightcurve and sufficient sensitivity to recover this structure in two images with different magnifications. The lower panel shows a more difficult, but likely characteristic scenario in which the burst is described by a single FRED-like pulse, at moderate signal-to-noise and the image(s) at lower magnification are only weakly detected. The latter scenario may have a different duration, while the weak signal will limit the ability to determine spectral and lightcurve parameters with confidence. Events pairs like this are present in the BATSE and {\em Fermi} catalogs \cite{}, but their physical relation remains unclear. }
	\label{lensing_cartoon}
\end{figure*}

There are several advantages of GRB lensing over the traditional routes to identifying lensed systems, and these, in turn, enable access to a series of different science questions. An ideal case for GRB lensing would involve an identical temporal and spectral pair of GRBs with a slight time and flux amplitude separation and is shown graphically in Figure~\ref{lensing_cartoon}. In this case many of the advantages listed below are present: 

\begin{itemize}
\item GRB detectors are by their nature wide field-of-view. Hence, a sensitive GRB detector can cover a significant fraction of the sky at any given moment. Lensing searches are not restricted to specific fields. Furthermore, they are sensitive to all time-delays, and not impacted by, for example, survey cadence or spatial resolution. 
\item GRB lightcurves are highly diverse so that, at least in principle, it is possible to identify a lensed GRB based on a lightcurve alone, potentially even with the very poor (several degree) error localisations provided by some GRB detectors. 
\item Because high-energy detectors provide intrinsic spectral resolution (unlike optical/IR imaging), GRBs can be readily cross-correlated based on their spectral {\em and} temporal properties to determine if bursts with morphologically similar lightcurves also have identical spectra as expected for achromatic lensing. 
\item The high luminosities of GRBs can be detected to high redshift, with the most distant GRBs occurring at $z>8$ \cite{tanvir09,cucchiaria11}. This is much more distant than detections of lensed supernovae (or indeed any supernovae) to date. The optical depth to lensing is maximized by these high redshifts \cite{smith23}.  
\item GRBs exhibit millisecond timescale variability, providing the capability to match two lightcurves with a remarkable degree of accuracy. In turn, this provides exquisite determination of the arrival time difference between two distinct images of a GRB. This enables precise measurements of, for example, cosmological parameters, but also the capability to probe to much shorter time delays and lower lens masses than possible with most lensed systems. If short-delay lenses are relatively more common as suggested by several studies \cite{mao92,grossman94} then GRB lensing would be a powerful route to uncover them. 

\end{itemize}

\section{The challenges of GRB lensing} 
It is also important to consider the challenges in identifying lensed GRBs. These are significant and some combination of these are responsible for the absence of any confirmed lensed GRBs to date. These challenges are listed below and some are shown graphically in Figure~\ref{lensing_cartoon}: 

\begin{itemize}
\item The intrinsic luminosity (and indeed flux or fluence) distributions for GRBs are extremely broad, spanning at least a factor of a million. It is not possible to pre-select bright GRBs as being those most likely to be lensed. 
\item While many GRBs exhibit complex lightcurves a large fraction are relatively simple and can be described by so-called Fast Rise Exponential Decay (FRED) pulse complexes. The most common number of such pulses in a GRB is one \cite{guidorzi24}. While the duration of this pulse is highly variable between GRBs, different magnifications for GRBs could yield different overall durations above some limiting detector sensitivity \cite{littlejohns13}. Hence, with poor localization obtaining a match by chance has non-zero probability. 

\item Most GRBs are only detected at modest signal-to-noise by the detectors in question. This impacts lensing detection in two ways. Firstly, the errors of the light curve and spectral parameters are substantial, increasing the possibilities of chance matches with physically unrelated bursts \cite{ahlgren20}. Secondly,  if we observe the most highly magnified event of a lensed GRB complex, the subsequent events could readily fall below the detection limits. 

\item The total number of GRBs detected since their first announcement \cite{klebesadel73} is around 10,000 \cite{burns23}. This is far smaller than the sample sizes used in searches for most other lensing events (although still, at present, substantially larger than the sample of gravitational wave events). While statistically, it is likely this sample does contain some lensed events given the optical depth to lensing at a median $z \sim 2$ \cite{smith23}, we have so far failed to identify one robustly. 

\item This lack of robust detections may arise partly
because of the poor localization in most GRB detectors. While pairs of similar GRBs 
have been detected we have not, to date, been able to robustly identify the lensing object and hence confirm that the burst is lensed. 
\end{itemize}

\section{Linking to multi-messenger lensing}

A particular appeal of GRB lensing is that it may be possible to use the lensed GRB to confirm a lensed gravitational wave signal. Most GRBs and the vast majority of gravitational wave detected compact object mergers currently have extremely large error boxes, and so the probability of chance coincidence of two apparently similar GW signals is non-zero, especially given the degeneracy between chirp-mass and distance that means gravitationally lensed GW events can also not be readily identified based purely on the waveform. This means that as with GRBs there are candidates for GW-lenses that cannot be confirmed. Indeed, a rather large fraction of the GW systems have been suggested by some authors to be lensed \cite{broadhurst1,broadhurst2,broadhurst3}, although such a large contribution is not expected. These challenges may be overcome with GRB signals, since while for either GRBs or GW events the individual probability of a chance spatial overlap of similar signals is non-zero, the probability of seeing both GW and GRB is vanishing small. To first order, the rate of GRB detections by {\em Fermi} is several per week, perhaps comparable to the GW detection rate, however, coincidence would only be considered with timing windows within a few seconds. The coincidence of a GRB and GW signal is already a significant event. The presence of two similar GRB signals simultaneous with two GW signals, all consistent with lensing, would be several orders of magnitude smaller still. Hence, the joint coincidence would immediately confirm the GW lensing signal. 

The difficulty in this scenario is that very few GW signals are expected to yield GRBs. In the first case, most GW merger signals arise from binary black hole (BBH) mergers. Although a controversial coincidence with a GRB was claimed for the first BBH identified \cite{connaughton2016}, it is generally not expected that these systems should yield luminous electromagnetic signatures outside rather special and unusual circumstances. Even for cases involving a neutron star, and for which matter remains outside the innermost stable orbit of the merged system, in most cases, we will not identify a GRB because they are beamed into a small fraction of the sky so that only a few percent will yield GRBs. This fraction may increase if some more isotropic component of emission is present as suggested by some authors, for example, X-ray emission from a nascent magnetar \cite{xue19,sun19,lin22}, but it seems likely this is a rare scenario. It is likely that we will need to wait until the third generation of gravitational wave detections (the Einstein Telescope and Cosmic Explorer) before there is a realistic chance of this scenario.

\section{A brief history of gravitational lensing in GRBs} 

\subsection{Lensing of prompt emission}
The first suggestions of lensing in GRBs stretch back to early papers advocating a cosmological origin. In particular, in the seminal paper making the case for a cosmological origin for GRBs Bohdan Paczy\`{n}ski argued that three, near identical bursts from the source B1900+14 were, in fact, a result of gravitational lensing \cite{pac86}. While the message of this paper -- that most GRBs lie at cosmological distances -- was undeniably correct, the lensing claim was ultimately spurious, as the outbursts of B1900+14 observed in 1979 were, in fact, the result of a magnetar outburst from a soft gamma-repeater in the Large Magellanic Cloud \cite{kouveliotou99}. In turn, this highlights the substantial challenges in identifying lensed GRBs -- in most cases, the sources are sufficiently poorly localized that the lens itself cannot be identified. 

The advent of the Burst and Transient Source Experiment (BATSE) greatly increased the number of detected GRBs, and in turn, the possibility that lensed GRB pairs would be observed. This led to the development of methods \cite{wambsganss93,nowak94} and expectations for lensing properties in the BATSE catalog, with a conclusion that the majority of lensed systems should have short delays, often of the order seconds to minutes \cite{mao92} although also extended to days and longer with lower probability \cite{grossman94}. Subsequently, multiple searches for identical GRBs have been conducted in BATSE data. While many of these have concluded with null results \cite{nemirof94, li14} there are cases worthy of further investigation. 

A particularly interesting case is the very short delay in the possible lensed GRB 950830 which shows two short, near identical outbursts with flux ratios of $\sim 1.5:1$ and a separation of only 0.4 s. This has been suggested to arise from lensing by an intermediate-mass black hole \cite{paynter21}. Such a result would be extremely important both as a lensed GRB, and because of the demonstration of GRBs as a route to probe extremely low-mass lenses. However, \cite{mukherjee24a} suggest that this burst does not show the expected achromatic behavior. Furthermore, it is challenging to disentangle intrinsic variability in a GRB, e.g. a GRB with two similar pulses from the same engine from the case in which a single pulsed GRB is lensed. Hence, while GRB 950830 is perhaps the best candidate for (prompt emission) lensed GRB to date, it remains unclear if it is truly the case. Indeed, in the case of very low-mass lenses, it may be impossible to robustly identify the lens, even in the case of precise positions. 

Further work has been undertaken with the large {\em Fermi sample}, with \cite{ahlgren20} identifying three possibly lensed GRBs, although it is also possible that all three are spurious. \cite{ahlgren20} also note that while the bursts appear similar, the substantial error bars on the light curves in many cases make simple comparisons of limited value in ascertaining lensing probabilities. 

It is notable that the majority of these GRBs are identified as being spatially coincident (with a large error box) but are widely separated in time. The time delays of several years are much longer than those seen in the majority of quasar or supernova lensed systems and perhaps indicate that they are more likely to be chance coincidence than physical associations.  Indeed, given the sensitivity of GRBs to short-duration lenses, and the expected higher density of low-mass to high-mass lens systems within the Universe (reflected in predictions for the delays for GRBs \cite{mao92,grossman94}), we should likely place more weight on lensing with shorter delay intervals compared to the longer ones. Indeed, simple trial factors would imply this, because the number of comparisons that one can perform (e.g., the number of bursts with consistent spatial locations) goes up linearly with the duration between the two candidate images. 

\subsection{Lensing of afterglows and associated supernovae or kilonovae}
In addition to the prompt emission, any GRB that is lensed will also exhibit a lensed afterglow (or more broadly it will be lensed at all wavelengths). In cases of short delays and hence small angular scales, we may not observe multiple images, and this may make the identification of the lensing signal difficult or impossible, even with Very Long Baseline Interferometry. In these cases, we may identify lensing in two ways. The first is that a well-sampled lightcurve should include the imprint of the burst at later times as it repeats. This was considered, for example by \cite{chen22} who investigated the appearance of lensed copies of the burst in the afterglow light curve. The advantage of this approach is that follow-up observations of GRBs are typically made with narrow field instruments that have much higher sensitivity than the $\gamma$-ray detectors themselves. As such, they may identify lensing signatures in bursts that are faint and so in which the prompt emission may not reach the necessary thresholds to re-trigger the $\gamma$-ray detectors. The challenge is that identifying the GRB itself is non-trivial because the later observations are made in a fundamentally different waveband than the prompt observations. This can be resolved if other features in the lightcurves can be identified as repeating within the afterglow. Although activity and flares in the light curves are common \cite{nousek06}, they are not typically repeating, and so such a repeat could be a strong indication of a lensed GRB. Indeed, such signatures have been considered in detail for structured jets \cite{gao22}. \cite{chen22} find one plausible event in the sample they consider, the case of GRB 130831A -- which has a $\sim500$s delay, however, the X-ray and $\gamma$-ray data does not support this lensing hypothesis. There is also one case -- GRB 020405 -- where a second transient is present close to a GRB afterglow in which a foreground absorber is present \cite{masetti03}. This has been suggested to be a lensed event \cite{rapoport12}, although is also plausibly an unrelated foreground supernova.

\section{Candidate lensed GRBs}\label{sec:candidates}

A summary of bursts that are suspected to be lensed is shown in Table~\ref{table_example}. These have been identified via a variety of techniques described above.

In most cases, the large error regions make any confirmation of lensing very difficult, although further bursts with near-identical structure and consistent spatial locations may strengthen the case (despite the very long delays). For bursts that are well localized there may still be possibilities of providing further evidence for (or against) lensing in the form of identifying plausible source and lens galaxies along the line of sight. For example, in the case of GRB 220627A, there are two similar bursts with a 1000s separation in time. There are also at least two absorption systems along the line of sight, perhaps enhancing lensing prospects. However, in this case, the bursts are not spectrally similar, and this would appear to rule out a lensing interpretation (although see section~\ref{sec7}). 

\begin{table}[!h]
\caption{Summary of claimed possible lensed GRB candidates}
\label{table_example}
\begin{tabular}{lllll}
\hline
GRB & Type & GRB-lc  & delay & Reference\\ 
\hline
950830  & prompt & short, 1 peak & 0.4s & \cite{paynter21,mukherjee24a} \\
090717 & prompt & long, 3 peaks & 50 s & \cite{Kalantari,mukherjee24a} \\
020405 & afterglow/SN & long, 4 peaks  & days-weeks& 
\cite{masetti03,rapoport12} \\
081122A & prompt &long, 2 peaks & 15 s & \cite{mukherjee24b} \\
081126A & prompt &long, 1 peak & 30 s& \cite{mukherjee24b} \\
100515A/130206B & prompt & long, 1 peak & 998 d & \cite{ahlgren20} \\
110517B & prompt &long, multi-peak & 15 s&  \cite{mukherjee24b} \\
140430B/161220B & prompt & long, 1 peak & 965 d & \cite{ahlgren20}\\
160718A/170606A & prompt & long, 1 peak & 323 d & \cite{ahlgren20} \\
060428B & afterglow/SN & long, multi-peak & unknown  & \cite{perley07} \\
130831A & Afterglow & long, 1 peak &   500 s & \cite{chen22} \\
210812A & prompt & long, 1 peak & 35s & \cite{veres21,mukherjee24b} \\
200716C & prompt & long, 3/4 peaks & 2.2s &  \cite{wang21,mukherjee24a} \\
220627A & prompt & long, multipeak & 1000s & \cite{roberts22} \\
\hline
\end{tabular}
\vspace*{-4pt}
\end{table}

\section{Chromatic effects in GRB lensing}\label{sec7} 

In searches for lensed GRBs the assumption is that the lensing is achromatic, and hence that lensed GRBs should be identical aside from their amplitude. This 
assumption is motivated because the act of lensing is achromatic -- the degree of lensing is independent of the photon wavelength. However, lensing leads to the observer viewing lines of sight from the source which would not have been observed in the non-lensed scenario. Hence, the lensing is only achromatic if the emission from the source is isotropic on 
the small angular scales relevant for lensing. For extended sources,  lensing can alter the apparent spectral energy distribution because of the differential magnification of different regions of the lensed source \cite{er14}. GRBs offer a related, but rather different problem. In particular, the 
GRB we observe is created in a relativistic jet moving towards the observer with a substantial bulk Lorentz factor which may approach $\Gamma \sim 1000$. The GRB emission is highly beamed due to this relativistic velocity and an observer can only view a causally connected region of the jet, along the line-of-sight, within an opening angle equivalent to $1 / \Gamma$. For observers with different lines of sight, the corresponding causal region of the jet and the observed emission may plausibly appear unique \cite{perna09}. 

\begin{figure*}
	\centering
		\includegraphics[scale=0.65,angle=0]{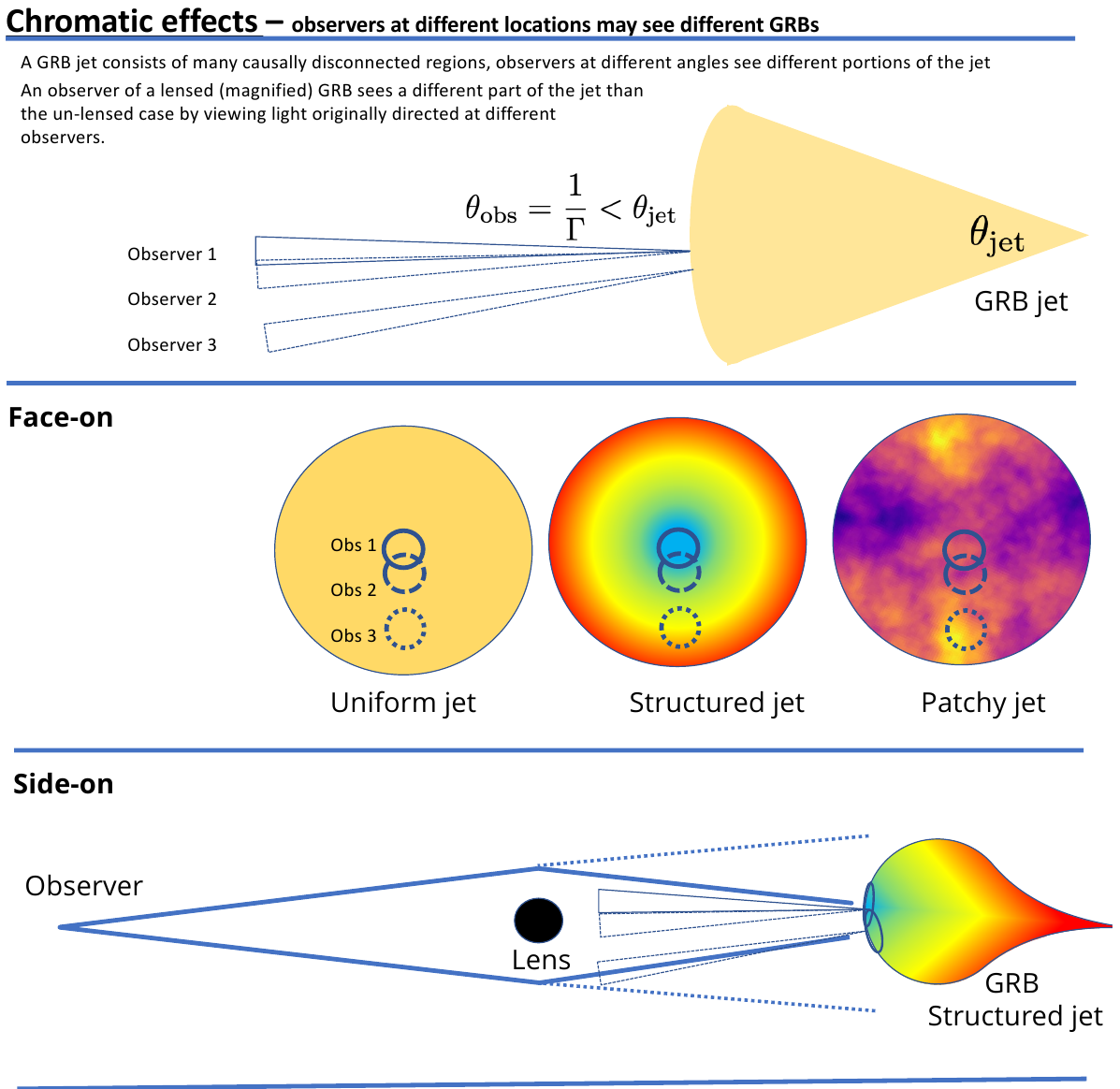}
	\caption{ Chromatic lensing for an anisotropic source. The top panel shows the general scenario for a non-lensed GRB. At early times the jet is moving at a high Lorentz factor and the narrowness of the observed region is due to relativistic rather than geometric beaming. This means multiple regions of the jet are not causally connected and may have different light curves depending on the jet structure. In a lensed scenario, we may then observe emitting regions that partially overlap, or are distinct from the non-lensed case, and we show how this may differ for various jet structures in the middle panel and lower panels. We also note that as well as the differing lines of sight, magnification also impacts the region of the jet observed and could give rise to further chromatic effects (e.g., \cite{granot01})} 
	\label{chromatic_lensing}
\end{figure*}

\begin{figure*}
	\centering
		\includegraphics[scale=0.35,angle=0]{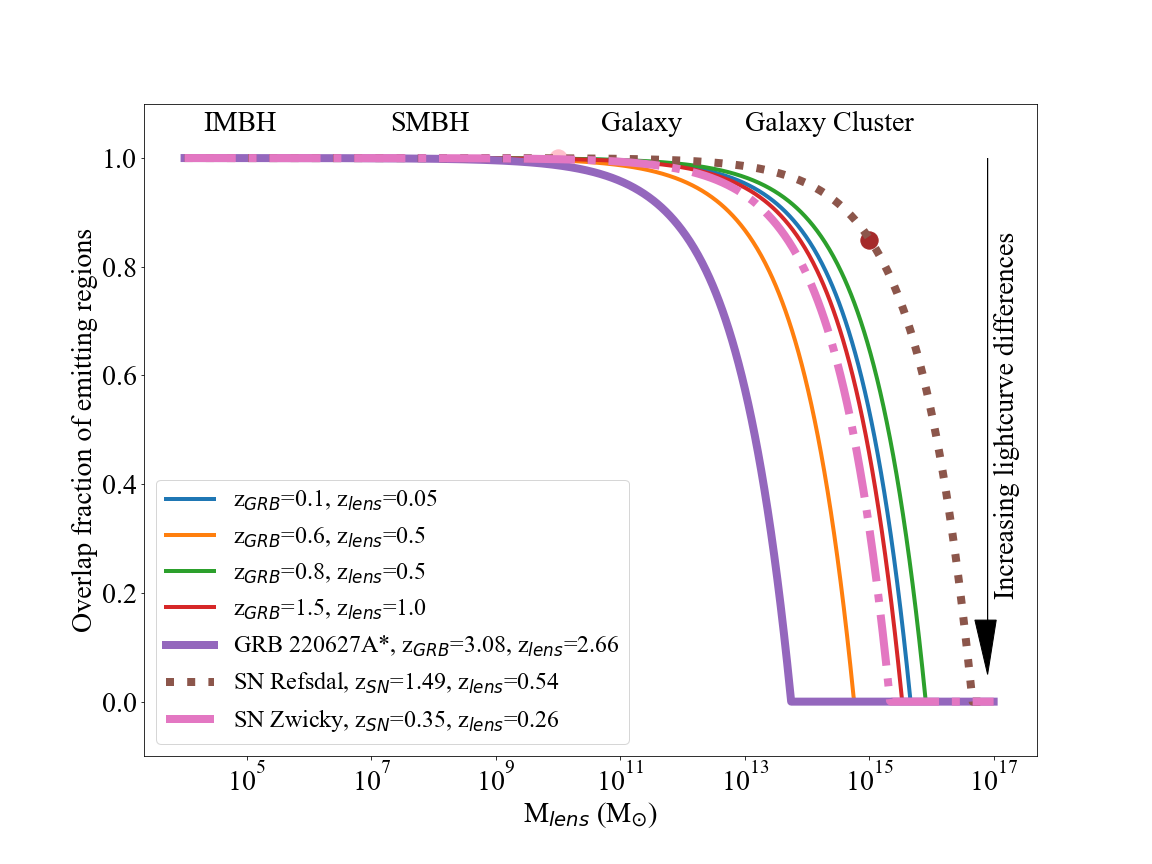}
	\caption{Scenarios in which chromatic lensing may be observed (see also \cite{perna09}. The figure shows the overlap region (one is complete overlap and 0 is no overlap) between causally disconnected regions of the jet for images separated by the Einstein radius for a given (simplified) lens. As can be seen, differences in the emitting region are generally only expected for very large masses, or in cases where the lens and source are close, which are only favourable geometries at very low redshift. We also mark on the figure the lens and source redshifts for strong gravitationally lensed supernovae, with the dots indicating the approximate mass of the actual lens. In the case of SN Refsdal, a GRB-like event may have exhibited lightcurve differences. We also include GRB 220627A which was suggested as a lensed GRB, but ruled out on the basis of non-identical lightcurves. This geometry could plausibly yield substantial differences even for galaxy mass lenses, but the time delay of only $\sim 1000$ s suggests a lower mass lens and that this impact is less important. However, magnification may still be important \cite{granot01}. }
	\label{lens_geom}
\end{figure*}

This problem is shown graphically in Figure~\ref{chromatic_lensing},
and has also been considered in more detail in \cite{perna09}, who were the first to point out the effect, and suggested that some lensed GRB pairs could simply be missed because the lensed image fell outside the jet entirely. Considering strong lensing in which the two widely separated (on the scale of arcseconds) images of the GRB are observed. In this scenario, we are observing different images of the source separated by roughly the Einstein radius ($\simeq \theta_{\rm E}$). At the lens, this separation can be tens to hundreds of kiloparsec, while the separation between the lens and the GRB is likely to be hundreds of Mpc to Gpc. Hence, for a 1/1000 ratio of image separation to lens-object distance, we would expect to observe an entirely different region of the jet. In practice, the geometry of the lens and object are likely to be more extreme, and the Lorentz factor may well not reach $\Gamma = 1000$.  However, even for a factor of 10 worse, we would expect to observe differences in causally connected regions of the jet order 10\%, which may yield comparably large changes in the light curves (especially for highly structured jets). We approximately show regions of interest in Figure~\ref{lens_geom}. Here we approximate a very simple point source lens with a background source that is closely aligned with the centre of the lens,
\begin{equation}
    \theta_E = \left({4 GM \over c^2} {D_{LS} \over D_L D_S} \right)^{1/2} \approx \left({M \over 10^{11.1} M_{\odot}}\right)^{1/2} \left({D_L D_S / D_{LS} \over \mathrm{Gpc}} \right)^{-1/2} \mathrm{\, arcsec}.
\end{equation}

Where the lens-source, lens, and source angular diameter distances are $D_{LS}$, $D_L$ and $D_S$ respectively.  We then define $R_E$ as being the physical Einstein radius at $D_L$, and determine the overlap in the emitting regions which becomes zero when the Einstein radius (in physical units at the lens) is larger than the opening angle $1 \over \Gamma$ (again in physical units) projected at the distance between the source and the lens (i.e. $\Gamma R_E > D_{LS}$). We then compare this overlap region for different masses of lens and a variety of lens geometries. Although the figure is only approximate and does not consider realistic lens models, it shows two generic features that are likely to remain true in most astrophysically realistic scenarios. Firstly, the effect is likely to occur only when either the lens mass is very high (on the scale of galaxy clusters, $\sim 10^{15}$ M$_{\odot}$), or when the lens and source become close to each other. Secondly, this implies that for anything but the most massive lenses in which the lens is equidistant between source and observer) are only likely in the local Universe (e.g. $z_{S} <0.2$). However, the optical depth of lensing here is such that it is unlikely to observe such a burst. To quantify this in plausible real cases we also include in Figure~\ref{lens_geom} lines showing the lens and source distances for SN Refsdal (cluster lens \cite{kelly15}) and SN Zwicky (galaxy lens \cite{goobar23}) as well as those possible for GRB 220627A (see below). Interestingly, had SN Refsdal been a GRB we may have expected some differences in the lightcurves due to the massive lens. No such differences would be expected for SN Zwicky. 

The second possible cause of chromatic behavior arises via magnification. Photons not originally bound for us traverse curved paths with the result that we observe patches of the jet outside the $1 /\Gamma$ causally connected region that we normally observe. A detailed discussion would be beyond the scope of this review, but the effect could be potentially important in some scenarios. In particular, such magnification has been considered in the case of afterglows which are subject to microlensing by lower mass compact objects \cite{granot01}.

The details of the effects of jet structure on the lightcurves of different lensed images are difficult to calculate because they depend on the details within the GRB jet. 
At one extreme there is the so-called patchy-shell model \cite{kumar2000, nakar2004, toma2006} in which different observers separated by angles $>1 / \Gamma$ see fundamentally different GRBs. At the other
extreme top-hat jets have essentially no variation across them and would yield identical GRBs to all observers (unless one is at the edge of the jet). More realistic and favored models today focus on structured jets in which the core of the jet has the most energy, and the energy per unit solid angle drops with increasing distance from the axis. Often this structure is assumed to be Gaussian \cite{zhang04}, although alternatives are also considered \cite{pascalli2015, lamb2021}. In a structured jet, the differences between the lightcurves would depend on this structure, and on the details of the forces between adjacent regions of the jet. There may be subtle or bulk differences in both lightcurve shape and spectrum. 

Whilst the effects discussed here largely concern the prompt emission, it is plausible that afterglow emission could show similar effects, despite  the Lorentz factor during the afterglow phase being typically lower than during the prompt emission. Hence the afterglow lightcurve shape may depend on viewing angle in this scenario, indeed lensed afterglows from structured jets have been considered by \cite{gao22}, as well as afterglow appearance in microlensed events \cite{granot01}.  

\begin{figure*}
	\centering
		\includegraphics[scale=0.25,angle=0]{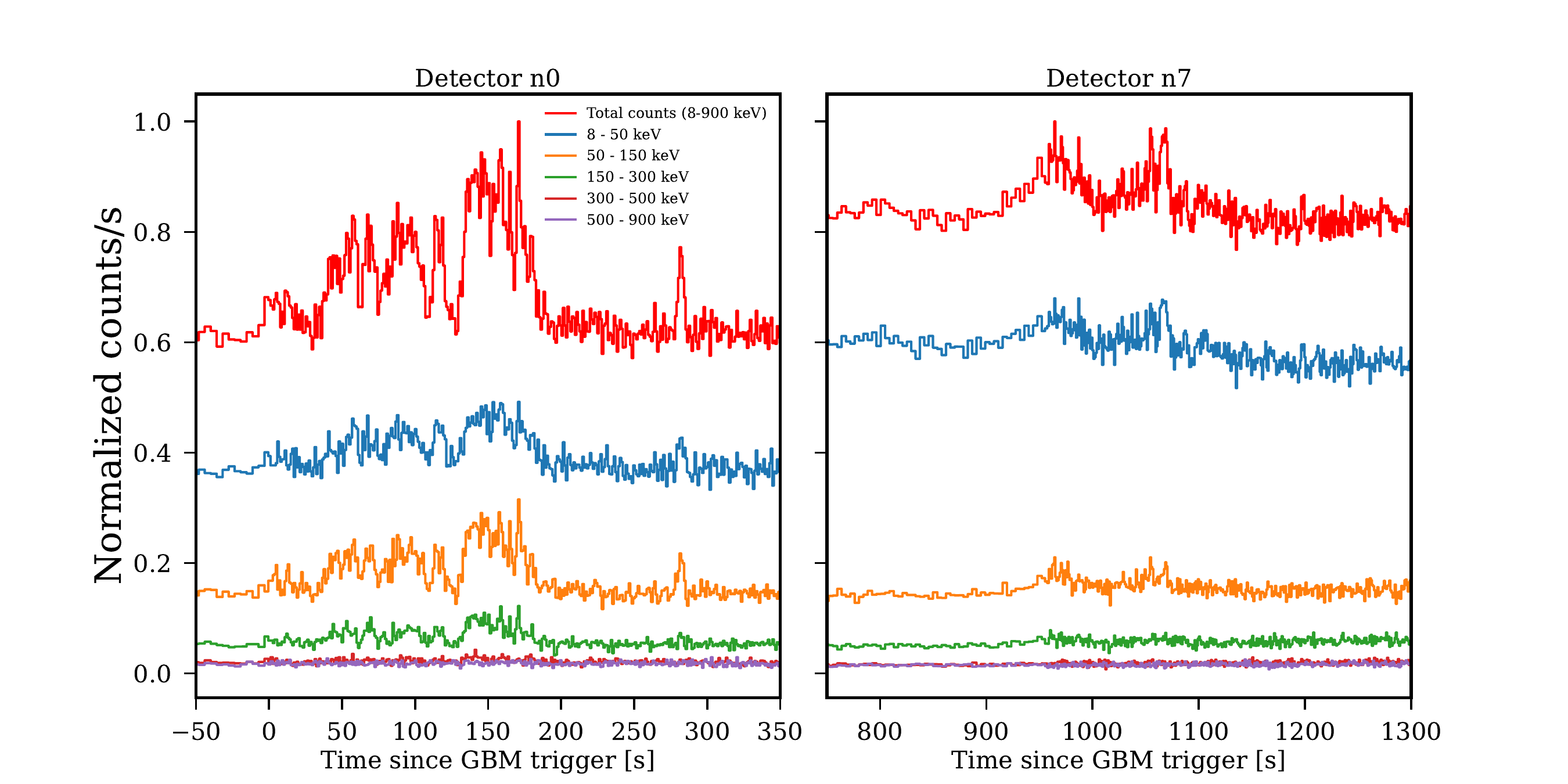}
	\caption{The lightcurve of the two 1000s separated bursts in GRB 220627A. The bursts have similar durations and shapes (with substantial errors, in particular on the second, fainter burst). In general, the preferred explanation is that GRB 220627A arises from an ultra-long GRB \cite{levan14} in which there are two broadly separated emission episodes \cite{huang22,dewet24}. However, if chromatic behavior is allowed because of the different magnification factors on the different images then lensing could remain a viable (if unlikely) explanation.  }
	\label{chromatic_lensing2}
\end{figure*}

Hence, it is not clear if lensed GRB emission will necessarily exhibit strong chromatic behavior, in most cases, it likely will not. However, these considerations also suggest that we should not automatically reject candidate lensed GRBs purely on the basis of chromatic variations, especially if these are small. This may have positive implications for the claims of lensing with sub-second delays in GRB~950830 \cite{paynter21}. It may also mean that other events, ruled out for lensing on the basis of chromatic behaviour could be reconsidered. One recent example of such a burst is GRB 220627A, the lightcurve of which is shown in Figure~\ref{chromatic_lensing2}. This burst exhibits two similar (but not identical) outbursts with a separation of 1000 \, s, and was considered a candidate for a lensing system \cite{roberts22}. It also has a strong intervening absorbing system which would imply a lens redshift of $z=2.665$ and a source redshift of $z=3.084$ (which is not optimal for lensing) \cite{huang22,dewet24}. A lensing origin for the GRB was disfavoured by \cite{huang22} on the basis of differing spectral shapes during the two peaks. However, given the substantial uncertainties on count rates, and the possibility of chromatic effects, this burst may still be a candidate lensed event. To further investigate this we also include this geometry in Figure~\ref{lens_geom}. Here, galaxy-scale lenses would have yielded significant differences because of the relative proximity of the lens and source. However, in this case, a time delay of only 1000s would be unlikely for a low-mass lens which would require extreme magnification for such a short time delay. For low-mass lenses, no significant lightcurve difference is expected. 

A further possibility raised by the discussion above is the prospect that lensing may allow us to probe GRB jet structure on small angular scales by observing slightly different angles {\em from the same GRB}. At present, we observe GRBs on axis, and in one case (GRB 170817A) relatively far from the jet-axis \cite{mooley18,lyman2018,margutti18,margutti21}. If lensed GRBs could be confirmed, for example via the detection of multiple images of the optical afterglow and the identification of the lens, then any variation in the lightcurves would immediately provide an indication of the degree of inhomogeneity, patchiness, or structure within the GRB jet. Indeed, in addition to macro-lensing, GRB jet structure may be well probed by microlensing observations via lenses of much lower mass that can resolve the jet, and may lead to chromatic variations \cite{granot01}. 

\section{Enhancing the recovering of future lensed GRBs} 
Clearly to date we have been unable to recover any lensed GRBs with high confidence, although calculations of the optical depth to lensing suggest that they should exist within the sample (Smith et al, this issue). Part of this challenge is that of the candidate events discussed in section~\ref{sec:candidates}, very few were suggested as lensed GRBs in real-time. This, in turn, prohibited follow-up observations that might have identified counterparts or plausible lenses. New developments can substantially enhance the prospects, and we suggest three likely routes through which this could be done: 
\begin{itemize}
    \item Cross correlating new GRB lightcurves with old ones in near real-time can identify lensed candidates on timescales when afterglow searches can be undertaken. 
    \item Much larger samples of events from new missions increase the possibility of lensing identification, especially if source locations are good.
    \item Improving location accuracy for poorly localised events (e.g. {\em Fermi}-GBM bursts by X-ray or optical afterglow localisations) greatly enhances the probability of successfully identifying lensed GRBs.
\end{itemize}

In the first scenario, cross-correlating new lightcurves and locations with old ones is something that could be done reasonably well on time scales of hours, enabling plausible candidates to be distributed on timescales on which counterparts may be found. Locating these counterparts, even in large error boxes is also more plausible now than in the past. Several counterparts to bursts discovered by 
{\em Fermi} have been found by ground-based follow-up surveys, most notably the Palomar Transient Factory and its successor the Zwicky Transient Factory \cite{singer13,singer15}. However, others have also been found by MASTER \cite{lipunov16}, GOTO \cite{belkin24} and MeerLICHT \cite{dewet24}, demonstrating that while not straightforward, such counterparts can be found. The combination of two poorly localized events can enhance the prospects because it is only necessary to search the overlapping confidence regions, say $\lesssim 10^2 -10^3\,\rm degree^2$. Furthermore, because such event pairs are likely to have long arrival time differences ($\Delta t\simeq1\,\rm day-1\,year$) they are likely to be associated with a massive galaxy/group/cluster-scale lens. Therefore identifying the lens should be possible even without finding multiple images of the afterglow. While to date optical searches for GRB afterglows have been substantially undertaken by small robotic telescopes, the advent of the Vera C. Rubin Observatory and her Target of Opportunity program would bring a significantly higher chance of success by probing quickly to much greater depths ($\rm AB\simeq24$), as long as candidate pairs of lensed GRB images are identified quickly (within a few hours). Given the likely small number of targets such a search could be done with no more than a single campaign per year tiling perhaps hundreds of square degrees \cite{rubinToO}.

The second route to identifying lensed GRBs is to create much larger samples of well-localized events. This sample has been growing at a rate of $\sim 100$ GRBs per year throughout the {\em Swift} lifetime, but new missions are providing a new route to enlarging these samples. There are two prime next-generation missions that may make substantial progress in this regard, the Einstein Probe (EP \cite{yuan18}), launched on 9 Jan 2024 and the Space and Variable Objects Monitor (SVOM), launched on 24 June 2024 \cite{svom11}. The Einstein Probe makes use of novel lobster eye technology to conduct a wide-field soft X-ray survey from 0.5-4 keV across 3600 sq. degrees, and the first discoveries indicate a high event rate and the ability to probe to high redshift \cite{gillanders24,levan24,lui24}. 
The WXT is potentially sensitive to a much lower flux than, for example, the {\em Swift}-BAT or {\em Fermi}-GBM. This raises the possibility of effectively running a forced photometry route at the location of previously identified GRBs and transients to search for any future activity. Indeed, since EP can also detect afterglow emission we may expect it to recover faint counter images of any lensed GRBs previously detected by {\em Swift}, or to use it to identify the counterparts of {\em Fermi} GRBs directly providing much smaller position uncertainties. The challenge will be in removing the (potentially many), for example, Galactic sources which may be expected to repeat. SVOM also has a Lobster-eye telescope in addition to a GRB monitor and optical telescope. It will operate in a route rather similar to {\em Swift}. Combined, these telescopes will be more sensitive than previous searches, will detect a larger number of sources, and enhance the fraction of the sky over which accurate positions can be obtained. 

The third route would be to use wide-field X-ray telescopes such as the EP WXT to provide precise positions for {\em Fermi} (or other poorly localized bursts). Since the WXT field of view is 3600 square degrees it could readily cover the error boxes of the majority of {\em Fermi} bursts, while its sensitivity would be sufficient to identify afterglow to a significant fraction. Indeed, to quantify this we can assume (to first order) that the {\em Fermi}-GBM bursts have similar afterglows to those seen with {\em Swift}-XRT and that the EP-WXT could respond to {\em Fermi} alerts on $\sim 10$ minute timescales. If this is the case, then it should recover the vast majority of afterglows (see, for example, Figure~3 in \cite{levan24b}).




\section{Conclusion}
Gravitationally lensed GRBs are yet to be unambiguously identified despite intensive but mostly archival efforts to locate them within the sample of $\sim 10,000$ GRBs observed by GRB missions over the past 50 years. The central difficulty in GRB lensing claims lies in the inability in most cases to directly identify the lens, and by the intrinsic range of properties exhibited by the GRBs themselves. We are now detecting GRBs at a higher rate than ever, and in particular, in the near future will have much larger detection rates of well-localised events. This enhances the probability of successfully identifying a lensed GRB based on the lightcurves alone. However, it remains a challenging task given the expected rates of such lensing, and enhancements in the rapid identification of lensing candidates would make substantial improvements in the probability of recognising, and subsequently confirming the first clearly lensed GRBs. 

\vskip6pt

\ack{We thank the Royal Society for hosting the meeting that enabled the discussion of these issues. }

\bibliographystyle{RS}







\end{document}